\documentclass[12pt,preprint]{aastex}
\usepackage[]{natbib,lscape,graphicx,graphics,verbatim}
\usepackage{lscape}
\usepackage{color}
\shorttitle{Nitrogen in the Pleiades and Hyades} 
\shortauthors{Miller et al.}

\begin{document}

\title{Nitrogen Abundances and the Distance Moduli of the Pleiades and Hyades} 

\author{Blake Miller, Jeremy R. King, Yu Chen}
\affil{Department of Physics and Astronomy, 118 Kinard Lab, Clemson University, Clemson, SC 29634-0978}
\email{blakem@clemson.edu{\ }jking2@clemson.edu{\ }yuc@clemson.edu}

\and
\author{Ann M. Boesgaard\altaffilmark{{\dagger}}}
\affil{Institute for Astronomy, 2680 Woodlawn Drive, Honolulu, HI 96822}
\email{boes@ifa.hawaii.edu}
\altaffiltext{{$\dagger$}}{Visiting Astronomer, W.M KECK Observatory, which is operated as a scientific 
partnership among the California Institute of Technology, the University of California and the National 
Aeronautics and Space Administration.  The Observatory was made possible by the generous financial support 
of the W.M. Keck Foundation.} 

\begin{abstract}

Recent reanalyses of {\it HIPPARCOS} parallax data confirm a previously noted discrepancy with the 
Pleiades distance modulus estimated from main-sequence fitting in the color-magnitude diagram. 
One proposed explanation of this distance modulus discrepancy is a Pleiades He abundance that is significantly 
larger than the Hyades value.   We suggest that, based on our theoretical and observational 
understanding of Galactic chemical evolution, nitrogen abundances may serve as a proxy for
helium abundances of disk stars.  Utilizing high-resolution near-UV Keck/HIRES spectroscopy, we 
determine N abundances in the Pleiades and Hyades dwarfs from NH features in the ${\lambda}3330$ region. 
While our Hyades N abundances show a modest ${\sim}0.2$ dex trend over a 800 K $T_{\rm eff}$ range, we find
the Pleiades N abundance (by number) is $0.13{\pm}0.05$ dex lower than in the Hyades for stars in a smaller
overlapping $T_{\rm eff}$ range around 6000 K; possible systematic errors in the lower Pleiades N abundance result are 
estimated to be at the ${\le}0.10$ dex level.  Our results indicate [N/Fe]${\sim}0$ for both the Pleiades
and Hyades, consistent with the ratios exhibited by local Galactic disk field stars in other studies.  
If N production is a reliable tracer of He production in the disk, then our results suggest the Pleiades
He abundance is no larger than that in the Hyades.  This finding is supported by the relative Pleiades-Hyades
C, O, and Fe abundances interpreted in the current context of Galactic chemical evolution, and is resistant to 
the effects on our derived N abundances of a He abundance difference like that needed to explain the Pleiades 
distance modulus discrepancy.  A physical explanation of the Pleiades distance modulus discrepancy
does not appear to be related to He abundance.

\end{abstract}


\section{Introduction} 

\subsection{The Pleiades Distance Modulus Problem}
A provocative early result of the ESA's {\it Hipparcos} parallax mission was a Pleiades distance modulus 
\citep{vLHR,vL} 0.30 mag smaller than that derived from evolutionary models via main-sequence fitting \citep{PSS}.  
Given that metal-poverty explains the small fraction of young nearby disk field dwarfs that are similarly subluminous 
in the {\it Hipparcos}-based color-magnitude diagram \citet{SKH}, \citet{PSS} suggested that the parallax versus 
main-sequence fitting discrepancy was caused by small (1 mas) systematic errors in the {\it Hipparcos} Pleiades 
parallaxes.  \citet{vvm} recomputed Pleiades parallaxes using the  {\it Hipparcos} intermediate astrometry data, 
finding a shift in the mean parallax that places the inferred distance modulus in substantial agreement with
the main-sequence fitting results.  \citet{RANT}, however, utilize Monte Carlo simulations of the {\it Hipparcos} cluster 
data, and find that no systematic biases or small angular scale effects are present in the Pleiades {\it Hipparcos\/} 
parallaxes.   

\citet{PSG} have cautioned against using lower main-sequence $(B-V)$ colors, which appear anomalous in the 
Pleiades \citep{St03}, in deriving distances from main-sequence fitting.  Excluding these data, \citet{PSG} find 
that various color-magnitude planes consistently yield Pleiades distances 10\% larger than parallaxes.  Distance
determinations of the Pleiades eclipsing binary HD 23642 \citep{SMS} and the Pleiades binary Atlas \cite{PSK04} 
are consistent with those determined from cluster main-sequence fitting, and inconsistent with the smaller distance 
implied by {\it Hipparcos}.   The HST/FGS-based parallaxes of 3 Pleiades dwarfs \citet{Sod05} are also in outstanding 
agreement with the main-sequence fitting-based results, but inconsistent with the {\it Hipparcos} result.

\citet{vL07a} and \citet{vL09} describe a new reduction, with improved treatment of small angular scale astrometric 
correlations noted by \citet{PSS}, of the {\it Hipparcos\/} mission astrometric data that has lead to significantly 
increased accuracy.  The result is a Pleiades distance modulus in accord with the original {\it Hipparcos} result.  
Indeed, \citet{vL09} cite 3 clusters (Pleiades, Blanco 1, NGC 2516) whose main-sequence stars appear similarly 
subluminous when compared to those in other clusters.

\subsection{Resolutions to the Pleiades Distance Modulus Problem: Metallicity}

\citet{G01} suggests that the discrepancy can be explained by a slightly sub-solar Pleiades metallicity, [m/H]$=-0.11$, 
indicated by Geneva photometry.  However, \citet{SN01} have used the metallicity-sensitive Stromgren $m_1$ 
index to select field stars having the same photometric metallicity as the Pleiades to form an empirical Pleiades  
ZAMS with which to conduct main-sequence fitting; \citet{SN01} deduce a Pleiades distance modulus in close 
agreement with previous fitting, but at odds with the {\it Hipparcos} parallaxes.  \citet{PSK03} 
also carry out empirical main-sequence fitting of the Pleiades using a sample of field stars of known 
metallicity and high accuracy {\it Hipparcos} parallaxes, and infer a surprisingly low Pleiades metallicity ([m/H]${\sim}-0.4$)
from a 2-color diagram.  Assuming this photometric metallicity, they find that the distance from main-sequence fitting 
is brought into agreement with the {\it Hipparcos} results.

The differences between the Geneva-, 2-color plane-, and Stromgen-based \citep[\rm{[Fe/H]}$=+0.08$;][]{Egg86} 
photometric Pleiades metallicity estimates, as well as the simultaneous claims that metallicities of both $-0.11$ and 
$-0.40$ shift the evolutionary model main sequences into agreement with the Hipparcos parallaxes, seem remarkable.  
The implication that a variety of independent modern high-resolution spectroscopic abundance analyses 
\citep{BF90,ccc88,Wild02,k00,fjs,sod09} have consistently overestimated a near-solar Pleiades metallicity also is remarkable.  

\subsection{Resolutions to the Pleiades Distance Modulus Problem: Helium}

\citet{vL09} posits that the dichotomy between the {\it Hipparcos}-based H-R diagrams of the young 
Pleiades, Blanco 1, and NGC 2516 clusters and older clusters like the Hyades and Praesepe is due to 
``age-dependent luminosity effects'' that are neither empirically calibrated out of the theoretical isochrones 
nor included in the underlying evolutionary models.  Such a luminosity difference between the Hyades/Praesepe 
and Pleiades main-sequences was inferred on the basis of Stromgren photometry long ago \citep{CP66}, 
questioned by \citet{Egg94}, but confirmed by \citet{JT95}.   \citet{vL09} note that this so-called ``Hyades 
anomaly'' has simply been forgotten, rediscovered from {\it Hipparcos} parallaxes, and relabeled as a 
Pleiades parallax anomaly.  

While several origins of a real luminosity difference between the Pleiades and Hyades main-sequences 
exist \citep{LR85}, a suggestion as old as the Hyades anomaly itself is inter-cluster He differences 
\citep{CS66} that yield inter-cluster luminosity differences.  \citet{PSS} note that a Pleiades 
He mass fraction of $Y=0.37$, compared to a solar value $Y_{\odot}=0.27-0.28$ \citep{BPW95}, could 
explain the 0.3 mag distance modulus discrepancy. Observational evidence for such large $Y$ values 
in young disk stars is muddy \citet{PSS}, and the determination of stellar He abundances is fraught 
with uncertainty and study-to-study differences.  A relative comparison of Pleiades He abundances with 
those in the Hyades, for which the {\it Hipparcos\/}-based and main-sequence fitting-based distance 
moduli agree, remains elusive:{\ \ }stellar He abundances are best-determined in B stars, which are 
absent in the Hyades due to its age.      

Chemical evolution models \citep[e.g.,][]{CP08} that map the coproduction of He, C, O, and Fe in 
the Galactic disk suggest He proxies could be used to examine the relative Pleiades-Hyades $Y$ values 
in the context of relative cluster C, O, and Fe abundances. While O abundances derived from high-excitation 
\ion{O}{1} lines exhibit unsettling large trends with $T_{\rm eff}$ \citep{SKHP,SHKKT} that 
complicate an assessment of relative cluster abundances, a reliable relative measure comes 
from the Schuler et al.~[\ion{O}{1}]-based determinations for 3 dwarfs in each cluster.  These data 
suggest {$\Delta$}[O/H]$=+0.01{\pm}0.05(4)$ (Pleiades$-$Hyades).  The \citet{CP08} chemical evolution 
models from \citet{CP08} yielding the largest $Y$ difference for this negligible O difference suggest 
${\Delta}Y=+0.004{\pm}0.02(2)$ (Pleiades$-$Hyades), significantly smaller than the ${\Delta}Y=+0.09$ 
prescribed by \cite{PSS} to resolve the Pleiades' distance modulus discrepancy.

High-excitation \ion{C}{1}-based C abundances for cluster stars with $T_{\rm eff}{\le}6500$ 
K, above which there exists evidence that dynamic transport mechanisms can alter the photospheric 
abundance \citep{GM08}, yield [C/H]$=+0.06{\pm}0.02$ and $-0.06{\pm}0.03$ for the Hyades and 
Pleiades \citep{FB90}. Interpreting these relative abundances with the \citet{CP08} models suggests 
$Y$(Pleiades) is no larger than $Y$(Hyades) at the ${\sim}3{\sigma}$ confidence level.  The difference 
(Pleiades $-$ Hyades) in mean cluster F-dwarf Fe abundance is ${\Delta}$[Fe/H]$=-0.16{\pm}0.03$ 
\citep{BF90}. The \citet{CP08} chemical evolution models suggest this corresponds to 
${\Delta}Y=-0.01{\pm}0.003$.   

\subsection{N as a He Proxy}

While these abundance comparisons are inconsistent with a He abundance difference able to
resolve the Pleiades distance modulus discrepancy, it may be that C, O, and Fe are not reliable 
tracers of He.  We suggest that N may serve as a more robust proxy for He, and that relative
cluster N abundances should thus be examined. 

Extant results suggest that [N/Fe]${\sim}0$ over a wide range of [Fe/H] 
\citep{Ecuvillon2004,Shi2002,Laird85,Carbon82}.  Such a primary nucleosynthetic production 
signature can not be produced by Galactic chemical evolution models that solely assume 
explosive nucleosynthesis in massive stars \citep{Timmes95}; rather, it is believed \citep{RV81,IT78} 
that N production occurs in low- and intermediate-mass stars
(LIMS; ${\sim}1{\ }{\rm M}_{\odot}{\le}M{\le}8{\ }{\rm M}_{\odot}$).  \citet{Carigi95} suggest 
that observations constrain the contribution of LIMS to N abundances in the solar neighbhorhood 
to $65-75\%$, with massive stars contributing most of rest and Type Ia supernovae contributing 
negligibly.  This is consistent with the estimate of \citet{WW95}, who ascribe only a quarter of 
solar N to that produced in massive stars. In contrast, O is the most abundant product of massive 
star explosive nucleosynthesis \citep{WW95}, over half of solar-metallicity Fe in the disk comes 
from explosive production \citep{Timmes95}, and only half of C in the solar neighborhood is 
produced by LIMS \citep{Carigi95}. 

While recent chemical evolution models identify non-explosive massive star wind yields as a 
source of uncertainty in Galactic He enrichment, the classic "road map" of \citet{WW95} identifies 
LIMS stars as the dominant nucleosynthetic He source (see their Table 19), and there is no 
question that most of mass shed by evolved LIMS stars on their way to becoming white dwarfs is 
in the form of H and He. The chemical evolution model and observational constraints from 
\citet{CP11} suggest that the LIMS contribution  to protosolar He (over and above the primordial 
Big Bang contribution) is ${\ge}50$\%. 

The nucleosynthetic results suggest that O and Fe are unlikely to be as robust proxies of disk 
He as are C and N, and that it is possible that N is a more robust proxy of He than is C in the 
Galactic disk.  The essential point is the importance of comparing the Pleiades-Hyades N abundances 
in considering the role of He in the Pleiades parallax anomaly.  We present such a comparison 
using self-consistently derived N abundances in solar-type dwarfs in the Pleiades and Hyades. 

\section{Observational Data}
 
High-resolution ($R{\sim}45,000$) spectroscopy of Pleiades and Hyades NH features near 3328 {\AA} 
was obtained over 3 observing runs in 1999, 2001, and 2002 using the Keck I 10m telescope and HIRES 
spectrograph.  The spectra are those used in the Be abundance studies of \citet{BK02} and \citet{BAK03}, 
who provide details concerning the observations and data reduction.  We utilize only a subset of these 
spectra here because the NH features become vanishingly weak and/or significantly blended in the 
hotter ($T_{\rm eff}{\gtrsim}6200-6300$ K) stars and/or similarly less amenable to abundance analysis 
for the stars with larger rotational velocities ($v{\ }{\sin}{i}{\gtrsim}15-20$ km s$^{-1}$).   Given the moderate 
trend in [N/H] abundance with $T_{\rm eff}$ we find for the (more numerous) cool Hyades stars, we 
restrict our attention in the Pleiades to stars with $T_{\rm eff}{\gtrsim}5950$ K, and also exclude stars 
for which binarity has been noted by others (e.g., \ion{H}{2} 739 and 761).   
 
Tables 1 and 2 list the Hyades and Pleiades stars analyzed here.  Hyades objects are listed with \citet{vb52} designations, 
while Pleiades stars are listed by their \citet{H47} identifications.   Examples of the spectra can be seen 
in Figure 1.  
\marginpar{Tab.~1}
\marginpar{Tab.~2}
\marginpar{Fig.~1} 

\section{Abundance Analysis and Results}

Stellar parameters taken from the Be abundance studies of \citet{BK02} and \citet{BAK03} 
were used to characterize LTE model atmospheres interpolated from the grids of 
Kurucz\footnote{http://kurucz.cfa.harvard.edu/grids.html}.  The linelist of the 3328 {\AA} 
region was compiled from atomic and molecular lines in the Kurucz 
database\footnote{http://kurucz.cfa.harvard.edu/linelists.html}, features in the Vienna Atomic 
Line Database \citep{Kupka2000}, and molecular lines from LIFBASE \citep{lifbase}. Our adopted 
NH dissociation energy adopted is 3.45 eV, intermediate to the canonical value of 3.47 eV 
\citep{Huber1979} and the determination of 3.40${\pm}0.03$ eV implied by experimental 
measures of various relevant quantities by \citet{Ervin1987}. Oscillator strengths
($gf$-values) were adjusted, typically by ${\le}0.2$ dex, to produce solar syntheses matching the Kurucz solar 
flux atlas \citep{Kur2005} assuming solar CNO (logarithmic number) abundances of 8.39, 7.78, 
and 8.69 \citep{AP01,Asp05}; all 3 abundances are prescribed because molecular equilibrium is included
in the syntheses. 

LTE synthetic spectra of varying N abundance were calculated using an updated 
version of the {\sf MOOG} package \citep{CS73} that includes updated bound-free opacity data 
important in the near-UV.  Input abundances for the syntheses are scaled to the solar 
values of \citet{AG89}; for CNO, the solar values given above are adopted.  We used scaling factors 
of [X/H]$=+0.13$ and $+0.00$ for the Hyades and Pleiades respectively\footnote{For both the 
Pleiades and Hyades, we assume [O/H]$=+0.14$ based upon the ${\lambda}6300$ [\ion{O}{1}]-based 
cluster dwarf results from \citet{SKHP} and \citet{SHKKT}.  For the Hyades, we assume 
[C/H]=$+0.15$ based upon initial abundances derived for three dwarfs from \ion{C}{1} and C$_2$ 
features \citep{SKT}.}.  Table 1 provides the $v$ ${\sin}i$ values adopted from the literature and used 
to smooth the syntheses in addition to a Gaussian representing instrumental broadening.  

We find a small (2-4\% of the continuum level) additional continuous veiling is needed to reproduce 
the depth of the strong non-NH features at  3327.9, 3328.3, 3328.9, 3329.5, 3329.9 {\AA} for our Hyades 
stars; this additional veiling, which might signal a slight deficiency in the bound-free opacity, also 
substantially improves the line-to-line scatter of the derived N abundances in a manner not mimiced by 
small plausible adjustments in the continuum normalization or smoothing (or both).  The additional veiling 
is added to our synthetic spectra by applying an additive constant prior to renormalization.  Syntheses and 
a comparison with observed spectra are shown in Figure 1.

The abundances in each star are determined from several NH features (or blended group of features) 
by fitting each individually\footnote{These features are located at 3325.88, 3326.39/3326.42, 3326.94, 3327.15,
3327.60, 3327.72, 3328.18, 3328.24, 3329.75, 3330.28, 3330.38, 3330.45, 3330.50, 3330.64, 3330.81, and 3330.92 
{\AA}}.  The scatter in these in a given star provides a combined measure of random 
measurement error and continuum fitting uncertainties in the derived abundances.   The abundance results 
are summarized in Tables 1 and 2.  The final three columns contain the mean N abundance 
(logarithmic by number, on the usual scale where log $N(H)=12.$), the number of features/regions 
used in determining the mean, and the standard deviation of the individual measurements.  The mean 
logarithmic number abundance of nitrogen for each star is plotted versus $T_{\rm eff}$ in 
Figure 2, which reveals a modest 0.2 dex trend in the Hyades dwarfs over the 5400-6200 $T_{\rm eff}$ 
range.  The non-parametric Spearman rank correlation coefficient is significant at $>99.9\%$ confidence 
level for the Hyades data.  
\marginpar{Fig.~2} 

We consider three possible sources of this trend.  First is a deficiency in continuous opacity-- 
whether truly continuous (e.g., bound-free) opacity or quasi-continuous opacity in the form of myriad 
very weak lines unnaccounted for in the linelist.  As noted above, we have guarded against such a 
deficiency by making small effective enhancements in the assumed veiling to reproduce the depths of
strong atomic lines and minimize the scatter in the N abundances derived from NH features of different 
strength. The second possibility is that our adopted NH dissociation energy is too low. However, 
observations of predissociation of electronic states of NH yield a robust upper limit of $D_{0}=3.47$ 
eV \citet{GL78}, a value insignificantly higher than our adopted value.  

The third possibility is an origin associated with the $T_{\rm eff}$-dependent abundance trends 
previously observed in Hyades dwarfs.  Figures 3, 5, and 10 of \citet{SHKKT} show a 0.5 dex monotonic 
increase in ${\Delta}$Fe$=$[\ion{Fe}{2}/H] $-$ [\ion{Fe}{1}/H], a 0.6 dex increase in \ion{O}{1}-based 
[O/H] values, and a 0.2 dex increase in [\ion{O}{1}]-based [O/H] values in Hyades dwarfs over the 
$T_{\rm eff}$ range 6000-4000 K.  Whether the trend we see for NH reflects an overdissociation akin 
to the apparent overexcitation/overionization these other abundances suggest, and what the physical 
origin of such effects are, remain unclear; a comparison of Li abundances derived from the 
${\lambda}6708$ resonance and ${\lambda}6104$ subordinate \ion{Li}{1} features in pre-main sequence 
stars, however, strongly suggests the action of enhanced near UV photoionization in cool very young 
stars \citep{Bubar2011}.  Regardless, we emphasize that the modest trends in N abundance are not 
surprising in the context of $T_{\rm eff}$-dependent trends previously seen in the Hyades dwarfs.

\section{Discussion}

Given the modest $T_{\rm eff}$ trend in the Hyades N abundances, we determine the relative 
Pleiades-Hyades cluster N difference over the same ${\sim}250$ K $T_{\rm eff}$ range spanned
by the four Pleiads.  For a given cluster, the standard deviation in the mean N abundance over 
this range, ${\pm}0.06$ dex for both clusters, empirically estimates internal measurement and relative 
stellar parameter uncertainties. This per star estimate is also that expected from the maximum 
$T_{\rm eff}$ uncertainties of \citet{BK02} and \citet{BAK03} and the typical mean measurement 
uncertainties calculated from the last 2 columns of Tables 1 and 2.  The unweighted mean cluster N abundances computed over the 5940-6180 K range are log $N$(N)$=7.78{\pm}0.03$ (uncertainty in the mean) and 
7.91${\pm}0.03$ for the Pleiades and Hyades, respectively. Weighting the individual abundances by 
the squared reciprocals of the individual uncertainties in Tables 1 and 2 yields indistinguishable 
mean abundances of log $N$(N)$=7.78{\pm}0.03$ (Pleiades) and 7.90${\pm}0.03$ (Hyades).  

The Pleiades $-$ Hyades difference is then ${\Delta}$log $N$(N)$=-0.13{\pm}0.05$, 
indicating that the Pleiades N abundance is smaller than that of the Hyades.  The mean cluster Fe 
abundances and their uncertainties from \citet{BF90} yield [N/Fe]$=0.00{\pm}0.04$ and 
$+0.03{\pm}0.04$ for the Hyades and Pleiades respectively, where the quoted errors reflect internal 
uncertainties in the mean.  We also consider three sources of systematic error.  First, we note that 
the solar N abundance was fixed for each feature by slight alterations in the $gf$ values when 
calibrating the line list. Thus, the solar-normalized abundances are derived self-consistently, and 
are $gf$-independent; indeed, altering the NH log $gf$ values by ${\pm}0.3$ dex, we find differential 
curve-of-growth effects to be ${\le}0.01$ dex.  Second, had we not employed the veiling corrections
for the Hyades stars, then the Hyades-Pleiades abundance difference would be {\it increased\/} by $0.02$ dex.  
Third, systematic differences in dereddened colors at the $0.02$ mag level for $(B-V)$ are possible;
in this case, the concomitant alteration to the abundances through the adopted $T_{\rm eff}$ values
is at the ${\pm}0.06$ dex level.  In sum, we gauge possible systematic errors in our finding of a Pleiades 
N abundance that is lower than that in the Hyades to be at the $0.06$ dex level. 

The simplest conclusions reached here, then, are that a) both the Hyades 
and Pleiades results are consistent with previous conclusions that [N/Fe]${\sim}0$ over a range in 
metallicity in the Galactic disk, and b) after accounting for possible systematic error, the Pleiades 
N abundance (by number) is the same as or ${\sim}25\%$ lower than that of the Hyades--thus providing 
no evidence that the Pleiades He abundance (by mass) is 40\% larger than that of the Hyades if indeed 
N production in the Galactic disk is a proxy for He production. 

Another possible systematic effect that must be considered, however, is the influence of He abundance 
on the derived N abundances themselves.  That is, we must ask if it is possible that the Pleiades He and N abundances might truly be enhanced relative to the Hyades, but our derived Pleiades N abundance is too low 
because we have not accounted for such a He enrichment in our analysis.  One effect of such a putative 
He enhancement would be on the Pleiades stellar parameters.  Equation 4 of \citet{Cast99} 
indicates that an enhancement of ${\Delta}Y=+0.10$ (that needed to explain the Pleiades-Hyades distance
modulus discrepancy) would lead to a 0.05 mag reduction in $(B-V)$ colors of solar metallicity $M_V=6$ 
Pleiades dwarfs.  The dwarfs we compare here are brighter, and an estimate of the color sensitivity for
them can be made using the reciprocal and reciprocity theorems to find 
$\left(\frac{\partial c}{\partial Y}\right)_T$ (where $c$ is $(B-V)$ color, $T$ is the effective 
temperature, and $Y$ is the He mass fractions) by taking: {\ } 
$\left(\frac{\partial M_V}{\partial Y}\right)_c$ from equation 3 of \citet{Cast99}, 
$\left(\frac{\partial T}{\partial M_V}\right)_c$ from Figure 4 of \citet{Cast99}, and 
$\left(\frac{\partial c}{\partial T}\right)_Y$ from the calibration of \citet{SH85}.  The result is 
the same as above, with a $+0.10$ increase in He mass fraction leading to a $(B-V)$ color bluer 
by 0.05 mag.  

The result of such a helium-induced color shift would be to overestimate the Pleiades $T_{\rm eff}$ 
values by ${\sim}170$ K and to underestimate the log $g$ values by ${\sim}0.02$ dex.  Compensating
for these parameter errors, including the effects on [Fe/H] and the feedback of metallicity on the 
derived N abundance, would lower our N abundances by $0.17-0.18$.  Thus, any such He-induced parameter
effects act in the opposite way needed to mask a proposed truly higher Pleiades N abundance. 

A second effect of a putative higher Pleiades He abundance is that on the (model) photospheric structure.  
We have rerun our analyses using ATLAS12 model atmospheres with the standard solar He abundance 
and He abundances enhanced 33\% and 75\% by number.  For stars with $T_{\rm eff}=4800$ K, the He-enhanced atmospheres yield N abundances lowered by 0.04 dex and 0.08 dex compared to the solar He atmospheres; 
for stars with $T_{\rm eff}=6000$ K, the reductions are 0.03 and 0.07 dex.  Just as for the parameter-based effects, 
He-induced atmospheric structure effects act in the opposite way needed to mask a proposed truly higher 
Pleiades N abundance.  

\section{Summary}

Our analysis of high-resolution and -S/N near-UV spectroscopy yields a Pleiades N number abundance that is 
the same as or up to $25{\pm}9\%$ lower than in the Hyades.  This result is consistent with previous abundance 
work suggesting that [N/Fe] ratios of local Galactic disk stars are solar over a range of [Fe/H].  
If, as we argue, N production serves as a reliable proxy for He production in the Galactic disk, then our 
results provide no evidence for a Pleiades He abundance larger than that of the Hyades.  This conclusion 
is consistent with those reached from the relative Pleiades-Hyades C, O, and Fe abundances 
in the context of our current understanding of Galactic chemical evolution. This conclusion is also robust 
against the effects of an unrealized but truly higher Pleiades He abundance on model atmospheric structure and 
our stellar parameters.  If the Pleiades distance modulus discrepancy and Hyades anomaly are not due 
to unrealized systematic parallax and photometric measurement errors, then our results suggests their physical 
explanation is not associated with He abundance.

\acknowledgments
This work was supported by NSF grants AST 02-39518 and AST 09-08342 to J.R.K., and AST 05-05899 to A.M.B.

{\it Facility:} \facility{KECK}.

\clearpage

%input tables here
%\input{tab1.txt}
\begin{deluxetable}{l c c c c c c c }
\tablewidth{0 pt}
\tablecaption{Hyades Atmospheric Parameters{\tablenotemark{a}} and Abundances}
\startdata
\hline
\hline
Star  & $T_{eff}$ & log $g$ & $\xi$          & $v \sin i$     & ${\langle}\log{N}({\rm N}){\rangle}$ & $N$ & $\sigma$ \\
(vB)  & (K)       & cgs     & (km $s^{-1}$)  & (km $s^{-1}$)  &                                      &     & (dex) \\ 
\hline
 9  &	5538 &	4.44 &	1.06 &	3.4\tablenotemark{d}  &	7.63 &	16 &	0.12\\
 10 & 	5982 &	4.39 &	1.27 &	6.2\tablenotemark{b}  &	7.86 &	15 &	0.10\\
 15 &	5729 &	4.42 &	1.15 &	5.4\tablenotemark{b}  &	7.82 &	16 &	0.10\\
 17 &	5598 &	4.43 &	1.10 &	4.5\tablenotemark{b}  &	7.78 &	16 &	0.09\\
 27 &	5535 &	4.44 &	1.04 &	4.9\tablenotemark{b}  &	7.77 &	16 &	0.08\\
 31 &	6071 &	4.39 &	1.30 &	10.0\tablenotemark{b} &	7.89 &	15 &	0.13\\
 59 &	6120 &	4.38 &	1.32 &	5.00\tablenotemark{c} &	7.90 &	14 &	0.09\\
 61 &   6260 &  4.35 &  1.41 &  20\tablenotemark{c}   & 7.88 &   5 &    0.11\\
 63 &	5822 &	4.41 &	1.19 &	7.20\tablenotemark{c} &	7.84 &	15 &	0.08\\
 64 &	5732 &	4.42 &	1.15 &	3.4\tablenotemark{b}  &	7.88 &	16 &	0.07\\
 65 &	6200 &	4.37 &	1.36 &	8.8\tablenotemark{b}  &	8.06 &	14 &	0.15\\
 69 &	5435 &	4.45 &	1.02 &	4.60\tablenotemark{c} &	7.73 &	16 &	0.09\\
 87 &	5445 &	4.45 &	1.04 &	4.0\tablenotemark{b}  &	7.72 &	16 &	0.08\\
 92 &	5451 &	4.45 &	1.02 &	3.8\tablenotemark{b}  &	7.74 &	16 &	0.09\\
 97 &	5814 &	4.41 &	1.19 &	5.4\tablenotemark{b}  & 7.78 &	16 &    0.12\\
 106 &	5690 &	4.42 &	1.14 &	3.4\tablenotemark{d}  & 7.79 &	16 &	0.09\\
 113 &	6139 &	4.38 &	1.33 &	5\tablenotemark{c}    & 7.99 &	15 &	0.14\\
 114 &	5509 &	4.45 &	1.04 &	7\tablenotemark{a}    &	7.74 &	16 &	0.07\\
\enddata
\label{abundance}
\tablenotetext{a}{\citet{BK02}} 
\tablenotetext{b}{\citet{2003ApJ...125..3185B}} 
\tablenotetext{c}{\citet{2000AcA....50..509G}} 
\tablenotetext{d}{Estimated as part of our analysis from non-Nitrogen features} 
\end{deluxetable}

\begin{deluxetable}{l c c c c c c c }
\tablewidth{0 pt}
\tablecaption{Pleiades Atmospheric Parameters{\tablenotemark{a}} and Abundances}
\startdata
\hline
\hline
Star         & $T_{eff}$ & log $g$ & $\xi$          & $v \sin i$     & ${\langle}\log{N}({\rm N}){\rangle}$ & $N$ & $\sigma$ \\
(\ion{H}{2}) & (K)       & cgs     & (km $s^{-1}$)  & (km $s^{-1}$)  &                                      &     & (dex) \\ 
\hline
948 &	5960 &	4.39 &	1.36 &	$<12$ &	7.83 &	16 &	0.08\\
1794 &	5940 &	4.39 &	1.35 &	12 &	7.72 &	11 &	0.12\\
1856 &	6150 &	4.37 &	1.54 &	16 &	7.75 &	7  &	0.05\\
3179 &	6180 &	4.37 &	1.56 &	$<7$ &	7.83 &	10  &	0.09\\
\enddata
\label{abund}
\tablenotetext{a}{\citet{BAK03}}
\end{deluxetable}

%input figures here

\begin{figure}
\includegraphics[width=3.5in]{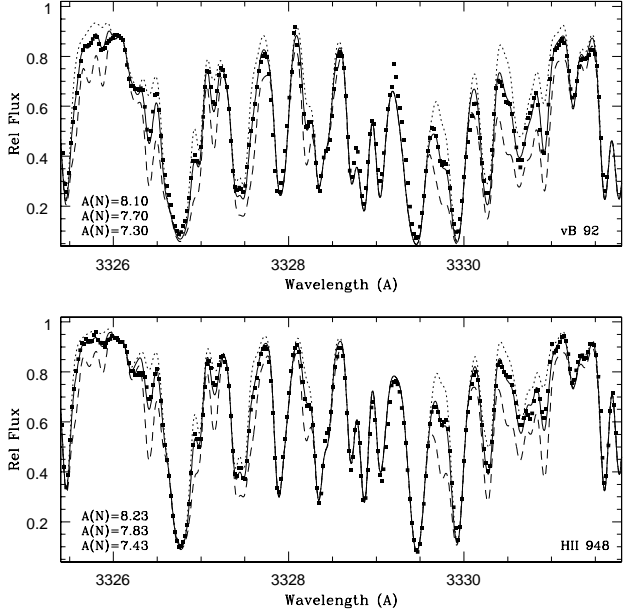}
\caption{Our observed spectra (solid points) of the Hyades dwarf vB 92 and the Pleiades dwarf \ion{H}{2} 948 are shown with synthetic spectra of varying N abundance;  $A$(N) indicates the logarithmic number abundance of nitrogen on the usual scale where that of hydrogen, $A$(H), is defined as 12.}
\label{fig:spectra}
\end{figure}

\begin{figure}
\includegraphics[width=3.5in]{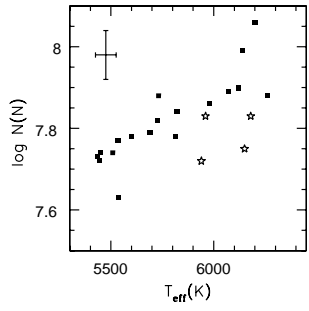}
\caption{The mean logarithmic number abundance of N for each star is plotted versus $T_{\rm eff}$ for our Hyades
(filled squares) and Pleiades (open stars) objects.  The error bar in the upper left shows the per star uncertainty estimated as described in the first paragraph of the Discussion.}
\label{fig:results}
\end{figure}

\end{document}